\documentclass[aps,pra,twocolumn,superscriptaddress]{revtex4}
\usepackage{amssymb}
\usepackage{amsmath}
\usepackage{graphicx}
\usepackage{epsfig}
\usepackage{subfigure}
\usepackage{color}

\begin{document}

\title{Classical correspondence of the exceptional points in the finite
non-Hermitian system}
\author{X. Z. Zhang}
\affiliation{College of Physics and Materials Science, Tianjin Normal University, Tianjin
300387, China}
\author{G. Zhang}
\email{zz@mail.nankai.edu.cn}
\affiliation{College of Physics and Materials Science, Tianjin Normal University, Tianjin
300387, China}
\author{Z. Song}
\affiliation{School of Physics, Nankai University, Tianjin 300071, China}

\begin{abstract}
We systematically study the topology of the exceptional point (EP) in the
finite non-Hermitian system. Based on the concrete form of the Berry connection,
we demonstrate that the exceptional line (EL), at which the eigenstates coalesce, can act as
a vortex filament. The direction of the EL can be identified by the corresponding Berry curvature. In this
context, such a correspondence makes the topology of the EL clear at a glance.
As an example, we apply this finding to the non-Hermitian Rice-Mele (RM)
model, the non-Hermiticity of which arises from the staggered on-site
complex potential. The boundary ELs are topological, but the non-boundary
ELs are not. Each non-boundary EL corresponds to two critical momenta that
make opposite contributions to the Berry connection. Therefore, the Berry
connection of the many-particle quantum state can have classical correspondence, which is
determined merely by the boundary ELs. Furthermore, the non-zero Berry
phase, which experiences a closed path in the parameter space, is dependent
on how the curve surrounds the boundary EL. This also provides an
alternative way to investigate the topology of the EP and its physical
correspondence in a finite non-Hermitian system.
\end{abstract}

\maketitle




\section{Introduction}

If a quantum system is conservative, a cyclic adiabatic change in a
parameter space, in which the wave function of the system follows
instantaneous eigenstates, can often give rise to surprising results such as
the emergence of gauge-invariant Berry phases \cite{Berry}. Of particular
interest is the case where eigenvalue degeneracies are enclosed within the
parameter loop. In the Hermitian regime, the corresponding Berry phase is
always the real number \cite{Xiao}. Over the past decades, there has been
intense interest in the complex Berry phase acquired during the adiabatic
evolution in the non-Hermitian systems \cite%
{Garrison,Dattoli,Ning,Moore,Gao,Pont,Massar,Ge,Whitney,Dehnavi,Nesterov,Cui,Liang,ZXZ}%
, which arises naturally in connection with various experiments. The Berry
phase generalized to non-Hermitian systems provides a geometrical
description of the quantum evolution of non-Hermitian systems \cite{Dehnavi}
and give the relationship between the geometric phase and the quantum phase
transition \cite{Nesterov,ZXZ}.

Another intriguing feature is the topology of the Berry phase, which has
gained considerable attention in the topological band theory\cite%
{Thouless,Kane,Moore2007,FuPRL,FuPRB,Schnyder,Kitaev2009,Bansil,Chiu,Armitage}%
. Topological insulators can be classified in the light of discrete
symmetries and dimensionality \cite{Chiu}. Topological properties of
one-dimensional (1D) systems with periodical boundary condition are
characterized by the so-called Zak phase, i.e., the Berry phase picked up by
a particle moving across the whole Brillouin zone \cite{Zak}. The Zak phase
is closely related to the electric polarization in solids and plays a key
role in the modern theory of insulators \cite{Xiao,Resta}. Most of those
studies focus on the Hermitian system. However, particle gain and loss are
generally present in natural systems that can be described by non-Hermitian
Hamiltonians \cite{Moiseyev}. This stimulates a growing interest in
topological properties of non-Hermitian Hamiltonians \cite%
{Hu,Esaki,TonyPRL,Leykam,Duan,Chong,Xiong}. Interestingly, non-Hermitian
systems compared with Hermitian ones exhibit highly nontrivial
characteristics. A fascinating example is non-Hermitian Hamiltonians at
exceptional points (EPs), where two (or more) eigenfunctions collapse into
one so the eigenspace no longer forms a complete basis. Most recently, the
topological nature of EPs in non-Hermitian Hamiltonians with chiral symmetry
has been recognized \cite{TonyPRL,Leykam} and the dynamical phenomena near
the EPs are being investigated both theoretically \cite%
{Uzdin,Berry2011,Berry2013,Gilary,Graefe,Moiseyev2014,Milburn} and
experimentally \cite{Moiseyev2016,Harris}.

Most of previous studies concentrated on the non-Hermitian many-particle systems with
continuous momentum, which requires the dimension of the system to be infinite.
The topology of such non-Hermitian systems can be characterized by the
modified Zak phase \cite{TonyPRL}, which is associated with the
bulk-boundary correspondence. For the finite non-Hermitian many-particle systems, however,
few studies have been done on the topological properties of EP.
Thus, a natural question to ask is whether the finite non-Hermitian many-particle system has obvious topological properties. These motivate us to study the topology of the EP in the finite non-Hermitian many-particle system. To this end, we systematically study the topology of the EPs
in a simple $2\times 2$ non-Hermitian system and then apply the conclusion to the
non-Hermitian Rice-Mele (RM) model that can be realized in experiment. Based on the analytical solution, we show that the Berry connection of the EL is connected to the curl field induced by the vortex
filament. This feature is protected by the chiral symmetry of the system. In
this point of view, the exceptional line (EL) directly corresponds to the
vortex filament. Therefore, the topology of the EL is straightforward.
Specifically speaking, when the system parameters are varied so
as to form a closed loop enclosing an EL in the parameter space, the
accumulated Berry phase is limited to an integer multiples of $\pi /2 $, which
depends on how the trajectory encircles the EL. This is topological since it
occurs only if the loop encloses the EL irrespective of its precise shape.
Furthermore, the Berry phase of the many-particle quantum state in the finite non-Hermitian system is merely determined by the boundary EL. When the dimension of the system increases to infinite, the Berry phase should be taken as a multiples of $\pi$ since the trajectory cannot surround a single EL. These also pave the way to investigate the topology of the EP and its physical
correspondence in a finite non-Hermitian system.

The remainder of the paper is organized as follows. In Section II, we
establish a bridge between the EL of non-Hermitian systems and the classical
electromagnetics. In Section III, we apply the classical correspondence to
investigate the topology of the non-Hermitian RM model. Finally, we give a
summary and discussion in Section IV.

\section{Classical correspondence of the topological invariant}

We start our analysis from a $2\times 2$ matrix consisting of pauli
operators,%
\begin{equation}
H=\alpha \sigma _{x}+\left( \beta +i\gamma \right) \sigma _{y},  \label{H}
\end{equation}%
where $\alpha $, $\beta $, and $\gamma $ are real numbers. The
nonhermiticity of this matrix arises from the complex coefficients of the
Pauli operator $\sigma _{y}$. This non-Hermitian Hamiltonian containing all
the information of the system not only can be employed to describe the
optical systems with gain or loss but also serve as the core matrix in the
non-Hermitian lattice system. It is worth pointing out that the Hamiltonian (%
\ref{H}) can be constructed through an arbitrary combination of the two
Pauli operators. For simplicity, we confine the discussion to the
Hamiltonians with $\sigma _{x}$ and $\sigma _{y} $. The following main
conclusion is still hold for any possible combinations of two Pauli
operators.

The eigenenergies of the system under consideration take the form $\pm
\varepsilon $, where
\begin{equation}
\varepsilon =\sqrt{\alpha ^{2}+\left( \beta +i\gamma \right) ^{2}}.
\end{equation}%
The two eigenenergies are complex owing to the existence of $i\gamma $. When
$\beta =0$, and $\gamma =\pm \alpha $, this system exhibits an EP at the
square-root branch point, where the two eigenstates coalesce and one of them
becomes defective. Here we refer to the two straight lines $\gamma =\pm
\alpha $ as ELs. The eigenstates of Hamiltonian $H$ can construct a complete
set of biorthogonal bases except the EP, which is associated with the
eigenstates of its Hermitian conjugate. For the considered system,
eigenstates $\left\vert \psi _{\pm }\right\rangle $ of $H$ and $\left\vert
\phi _{\pm }\right\rangle $ of $H^{\dag }$ form the biorthogonal bases and
are explicitly expressed as
\begin{equation}
\left\vert \psi _{\pm }\right\rangle =\frac{1}{\sqrt{2}\varepsilon }\left(
\begin{array}{c}
\alpha -i\beta +\gamma \\
\pm \varepsilon%
\end{array}%
\right) ,\text{ }\left\vert \phi _{\pm }\right\rangle =\frac{1}{\sqrt{2}%
\varepsilon ^{\ast }}\left(
\begin{array}{c}
\alpha -i\beta -\gamma \\
\pm \varepsilon ^{\ast }%
\end{array}%
\right) .
\end{equation}%
Now we focus on the symmetry of the system. The chiral symmetry of the
system admits that $\left\{ \sigma _{z},H\right\} =0$. So if $H$ has an
eigenstate $\left\vert \psi _{+}\right\rangle $ with eigenenergy $%
\varepsilon $, then $\sigma _{z}\left\vert \psi _{+}\right\rangle $ is also
an eigenstate with eigenenergy $-\varepsilon $.

\begin{figure}[tbp]
\centering
\includegraphics[ bb=150 340 470 630, width=0.38\textwidth, clip]{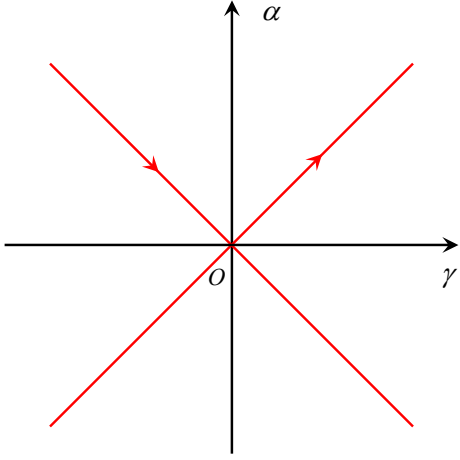}
\caption{(Color online) Schematic illustration of the classical
correspondence of the ELs. The two straight red lines denote the ELs of the
non-Hermitian system $H $, which can act as the vortex filaments owing to
the Eq. (\protect\ref{eqv}).}
\label{fig1}
\end{figure}

To capture the basic property of the dynamics around the EP, we start with
an adiabatic evolution, in which an initial eigenstate evolves into the
instantaneous eigenstate of the time-dependent Hamiltonian. For the sake of
simplicity, we assume that the Hamiltonian (\ref{H}) is a periodic function
of system parameters $\left\{ \beta ,\gamma ,\alpha \right\} $ that vary
with time $t$. Considering the cyclic time-dependent Hamiltonian $H\left(
\beta \left( t\right) ,\gamma \left( t\right) ,\alpha \left( t\right)
\right) $, it will return back to $H\left( \beta \left( 0\right) ,\gamma
\left( 0\right) ,\alpha \left( 0\right) \right) $ if all the parameters
change adiabatically to the starting point after time $\tau $. The adiabatic
evolution of the initial eigenstate $\left\vert \psi _{\pm }\right\rangle $,
which follows a closed path $C$ in the $\beta -\gamma -\alpha $ parameter
space, generates a Berry phase as%
\begin{equation}
\gamma _{B}=\int_{C}\overrightarrow{\mathcal{A}}\cdot \text{d}%
\overrightarrow{l}\text{,}
\end{equation}%
where $\overrightarrow{l}$ parametrizes the cyclic adiabatic process and the
Berry connection $\overrightarrow{\mathcal{A}}=\left( \mathcal{A}_{\beta },%
\text{ }\mathcal{A}_{\gamma },\text{ }\mathcal{A}_{\alpha }\right) $ can be
expressed as
\begin{equation}
\mathcal{A}_{g}=i\left\langle \phi _{\pm }\left( t\right) \right\vert \frac{%
\partial }{\partial g}\left\vert \psi _{\pm }\left( t\right) \right\rangle ,
\end{equation}%
where $\left\vert \psi _{\pm }\left( t\right) \right\rangle $ and $%
\left\vert \phi _{\pm }\left( t\right) \right\rangle $ represent the
instantaneous eigenstates of the Hamiltonian $H\left( \beta \left( t\right)
,\gamma \left( t\right) ,\alpha \left( t\right) \right) $ and $H^{\dag
}\left( \beta \left( t\right) ,\gamma \left( t\right) ,\alpha \left(
t\right) \right) $, respectively. $g=\beta $, $\gamma $, $\alpha $ denote
the three directions in the parameter space. It is worth pointing out that
the chiral symmetry give rise to the identical Berry connections for the
eigenstates $\left\vert \psi _{\pm }\left( t\right) \right\rangle $, namely
leads to $\mathcal{A}_{g}=\mathcal{A}_{g}^{\pm }$. After straightforward
algebras, the Berry connection can be expressed as
\begin{equation}
\overrightarrow{\mathcal{A}}=\left( \alpha ,\text{ }i\alpha ,\text{ }-\beta
-i\gamma \right) /\mathcal{D}
\end{equation}%
with $\mathcal{D}=2\varepsilon ^{2}$. To demonstrate the physical
correspondence of the Berry connection, we first assume that the EL can act
as a vortex filament. The corresponding curl field according to the
Bio-Savart law can be written as%
\begin{eqnarray}
\overrightarrow{A}_{\pm } &=&\left( -\gamma \pm \alpha ,\text{ }\beta ,\text{
}\mp \beta \right) /4D_{\mp }, \\
D_{\pm } &=&\left( \gamma \pm \alpha \right) ^{2}+\beta ^{2},
\end{eqnarray}%
where $\overrightarrow{A}_{\pm }$ represent the curl fields induced by the
two stright vortex filaments locating at $\gamma =\pm \alpha $,
respectively. Interestingly, the Berry connection and the curl field induced
by the vortex filaments can be connected through the gauge transformation%
\begin{equation}
\overrightarrow{\mathcal{A}}=\sum_{\sigma =\pm }\sigma \overrightarrow{A}%
_{\sigma }+\frac{i}{8}\ln \nabla \left( \frac{D_{-}}{D_{+}}\right) .
\label{eqv}
\end{equation}%
Here $\nabla $ is the nabla operator%
\begin{equation}
\nabla =\left( \overrightarrow{e}_{\beta }\frac{\partial }{\partial \beta },%
\text{ }\overrightarrow{e}_{\gamma }\frac{\partial }{\partial \gamma },\text{
}\overrightarrow{e}_{\alpha }\frac{\partial }{\partial \alpha }\right) .
\end{equation}%
In a closed path, the second term of the Eq. (\ref{eqv}) contributes nothing
to the Berry phase. Consequently, we deem the EL as a vortex filament and
sketch its direction in Fig. \ref{fig1}. The quantity $\gamma _{B}$ is
quantized in terms of classical physics and can be treated as a topological
invariant to characterize EL in the parameter space. The value of $\gamma
_{B}$ depends on the loop $C$ in the $\beta -\gamma -\alpha $ parameter
space: $\gamma _{B}$ is non-zero if the loop encircles an EL, while must be
zero if the loop does not encircle the EL. Such a correspondence establishes
a bridge between the EL of non-Hermitian systems and the vortex filament of
classical physics.

Here we point out that the aforementioned topological invariant of the
system may relate to the chiral symmetry, and this association does not rely
on the system parameters. $\left\{ \beta ,\text{ }\gamma ,\text{ }\alpha
\right\} $. In fact, if one consider the following Hamiltonian%
\begin{equation}
H^{\prime }=\mathcal{U}H\mathcal{U}^{-1},
\end{equation}%
where $\mathcal{U=}e^{-i\left( \overrightarrow{n}\cdot \overrightarrow{%
\sigma }\right) \theta /2}$ is a spin rotation operator and is independent
of the system parameters $\left\{ \beta ,\gamma ,\alpha \right\} $. Even
though the new Hamiltonian $H^{\prime }$ consists of three Pauli operators,
its eigenenergies are identical with those of $H$. This also indicates that
the two systems possess the same EL in the parameter space. Moreover, we can
redefine a chiral symmetric operator $\sigma _{z}^{\prime }=\mathcal{U}%
\sigma _{z}\mathcal{U}^{-1}$ to anticommute with $H^{\prime
}$, i.e, $\left\{ \sigma _{z}^{\prime },H^{\prime }\right\} =0$. Note that
the new chiral operator is also independent of the system parameters $%
\left\{ \beta ,\gamma ,\alpha \right\} $, which guarantees the identical
Berry connection of the two systems%
\begin{equation}
\mathcal{A}_{g}^{\prime }=i\left\langle \phi _{\pm }\left( t\right)
\right\vert \mathcal{U}^{-1}\frac{\partial }{\partial g}\mathcal{U}%
\left\vert \psi _{\pm }\left( t\right) \right\rangle =\mathcal{A}_{g}.
\end{equation}%
Thus all the conclusion about the considered system $H$ can be applied to $%
H^{\prime }$. In the following section we will apply this correspondence to
a finite non-Hermitian lattice system.

\begin{figure}[tbp]
\centering
\includegraphics[ bb=140 400 460 650, width=0.42\textwidth, clip]{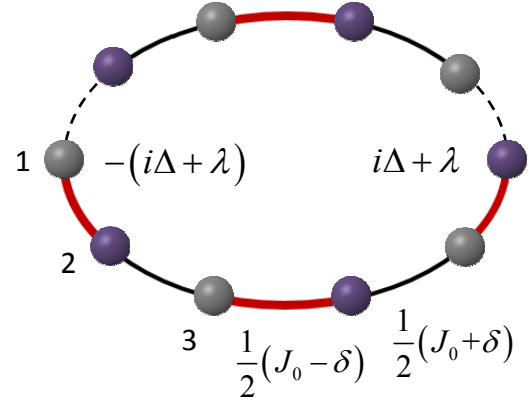}
\caption{(Color online) Schematic illustration of extended RM model with
complex on-site staggered potential. The red and black lines represent the
non-uniform hopping amplitudes. The purple and gray circles denote the
complex on-site staggered potentials with opposite values, respectively. The
system reduced to the standard Hermitian RM model when $\Delta =0$. Note
that the considered non-Hermitian system possesses the periodical boundary
condition, i.e., $c_{j}\equiv c_{j+2N}$, which is crucial to diagonalize the
Hamiltonian (\protect\ref{H_SSH}).}
\label{fig2}
\end{figure}


\section{Classical correspondence in a finte non-Hermitian lattice system}

\subsection{Model Hamiltonian and solution}

As an example, we consider a non-Hermitian extended RM model, which can be
described by the following Hamiltonian
\begin{eqnarray}
\mathcal{H} &=&\frac{1}{2}\sum_{j=1}^{N}\left[ \left( J_{0}-\delta \right)
c_{2j-1}^{\dagger }c_{2j}+\left( J_{0}+\delta \right) c_{2j}^{\dagger
}c_{2j+1}+\text{H.c.}\right]  \notag \\
&&+\left( i\Delta +\lambda \right) \sum_{j=1}^{2N}\left( -1\right)
^{j}c_{j}^{\dagger }c_{j}\text{,}  \label{H_SSH}
\end{eqnarray}%
the non-Hermiticity of which arises from the on-site staggered complex
potential $\left( i\Delta +\lambda \right) \sum_{j=1}^{2N}\left( -1\right)
^{j}c_{j}^{\dagger }c_{j}$. The system possesses a $2N$-site lattice, where $%
c_{j}$ is the annihilation operator of a fermion on site $j$. The nominal
tunneling strength $J_{0}$ is staggered by $\delta $. For clarity, we sketch
the structure of the system in Fig. \ref{fig2}. The origin Hermitian
Hamiltonian with $\Delta =0$ can be realized with controlled defects using a
system of attractive ultracold fermions \cite{Chin,Strohmaier,Hacke} in a
simple shaken one-dimensional optical lattice. Furthermore, the
non-Hermitian version can be achieved in a zigzag array of optical
waveguides with alternating optical gain and loss \cite{Longhi}. It is worth
mentioning that this non-Hermitian Hamiltonian can be also utilized to
control the wavepacket dynamics \cite{HWH,Lin}. Now we consider the
periodical boundary condition, that is $c_{j}\equiv c_{j+2N}$, to obtain the
exact solution.

\begin{figure*}[tbp]
\centering
\includegraphics[bb=0 0 408 311, width=0.4\textwidth, clip]{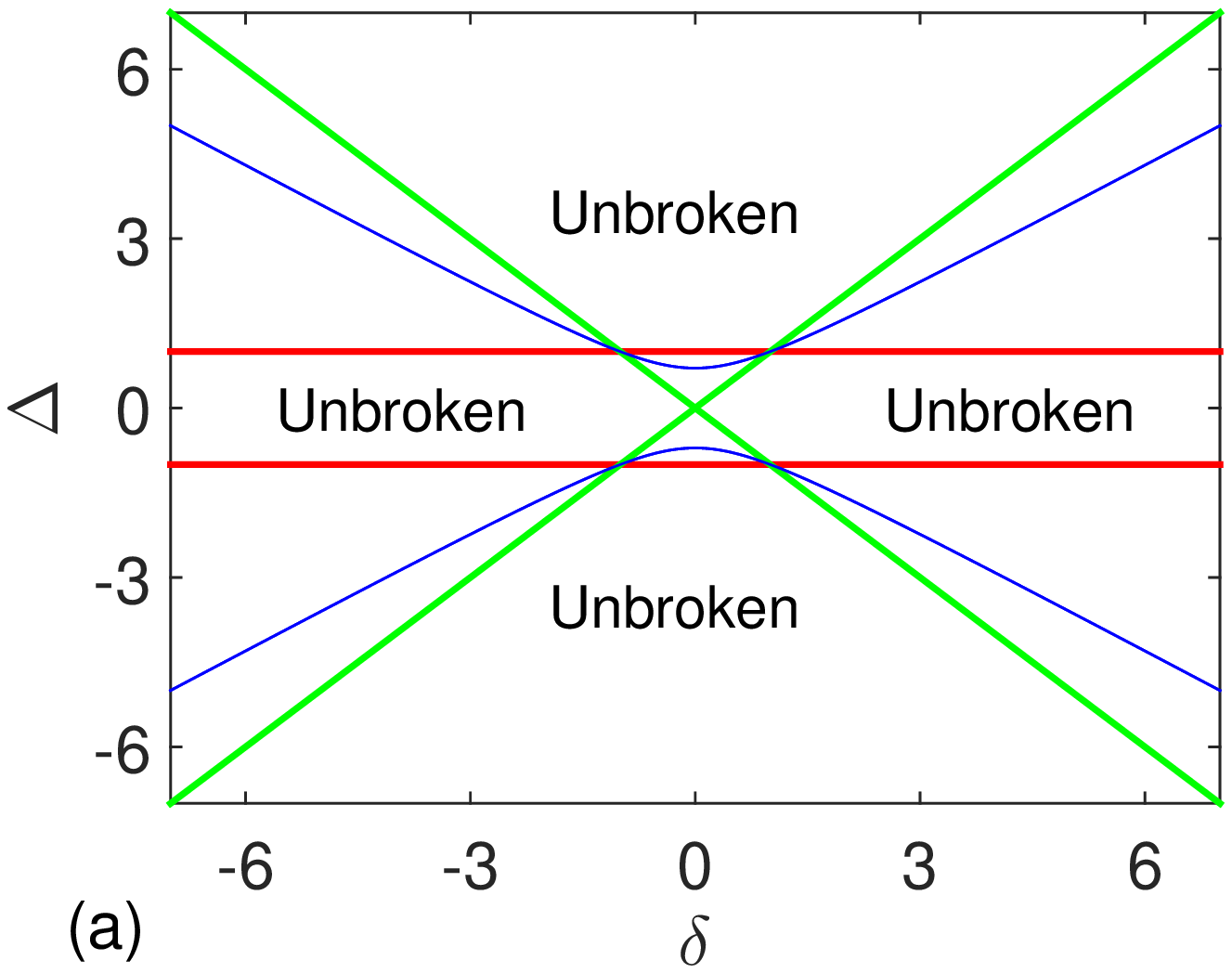} %
\includegraphics[bb=0 0 408 311, width=0.4\textwidth, clip]{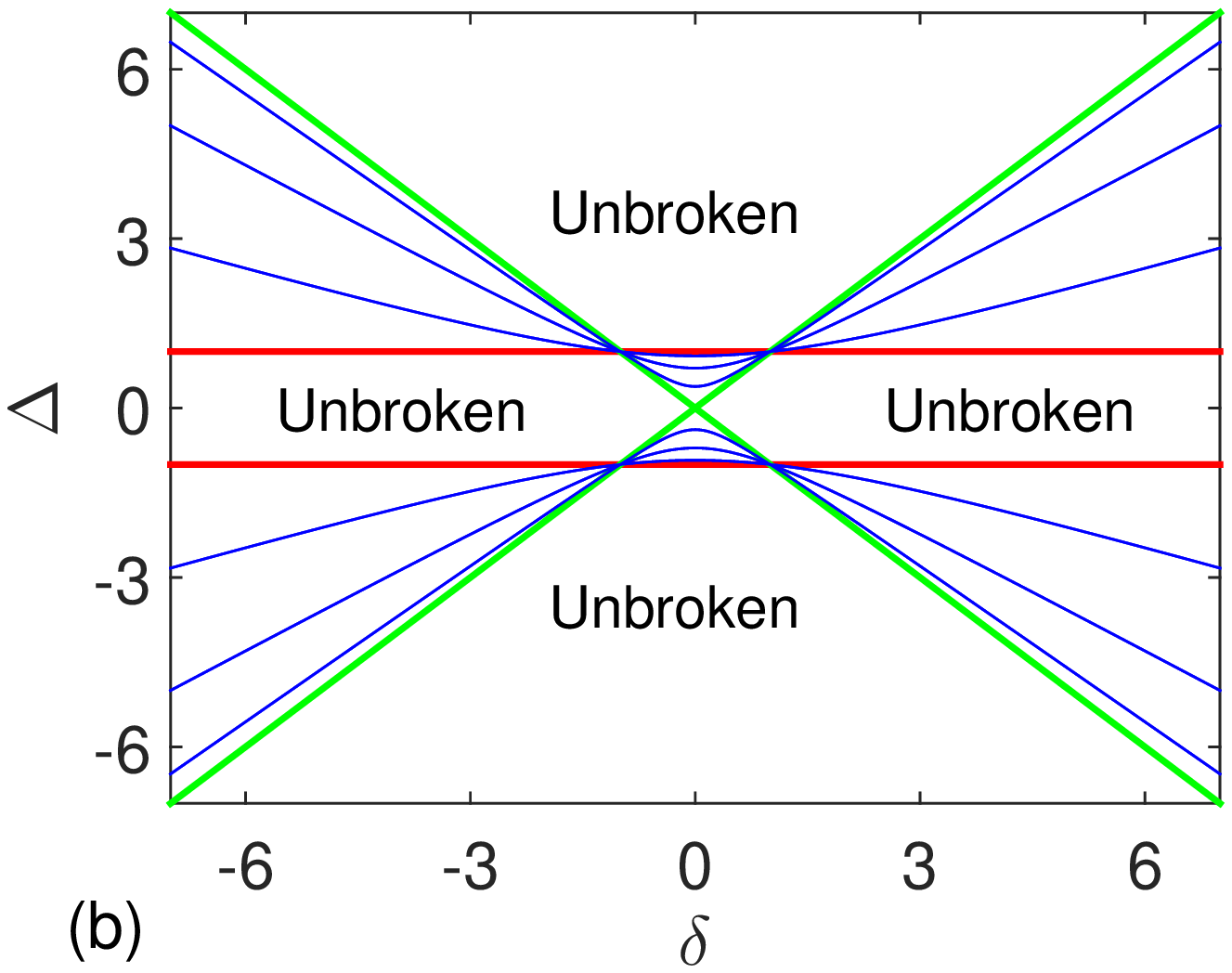} %
\includegraphics[bb=0 0 408 311, width=0.4\textwidth, clip]{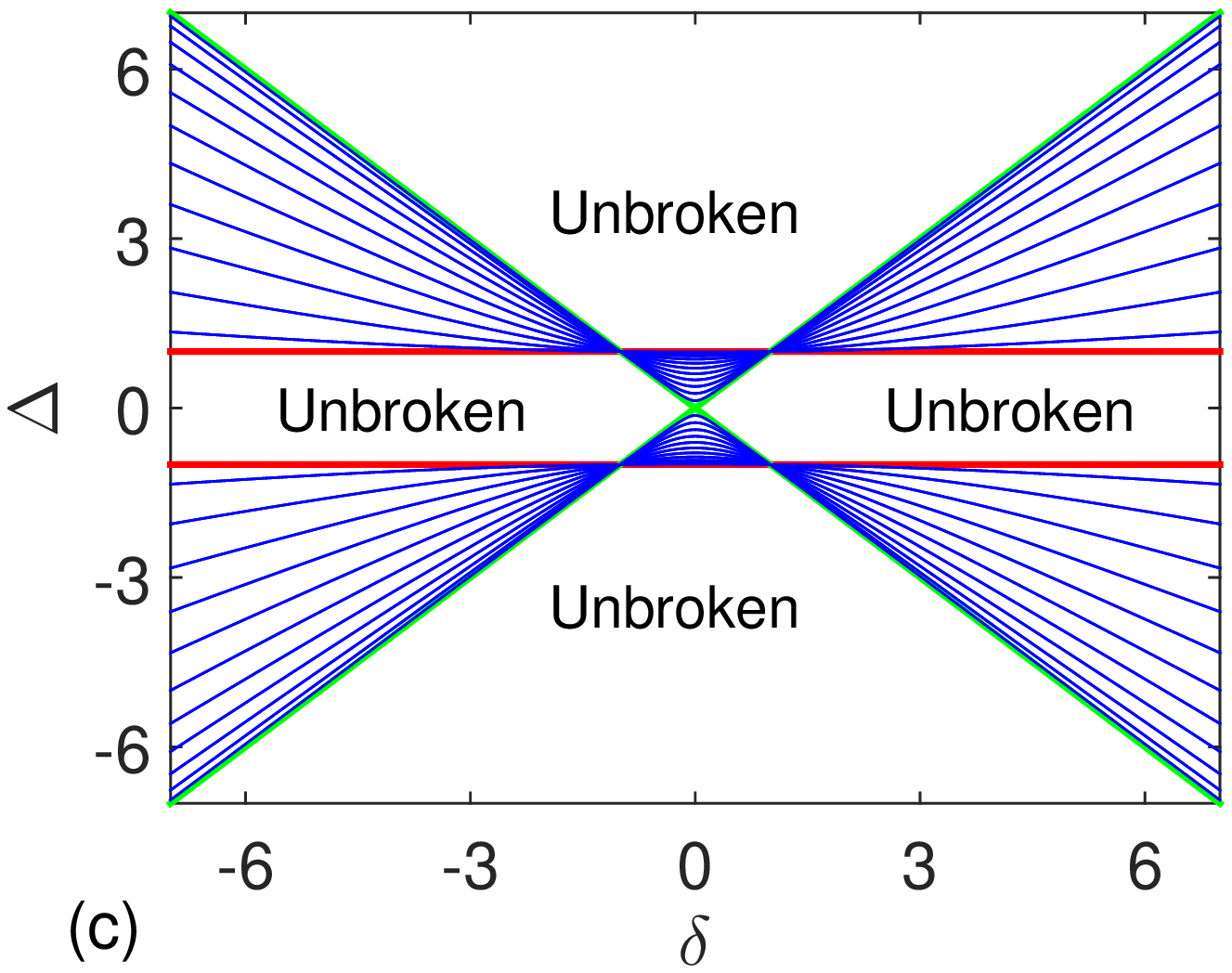} %
\includegraphics[bb=0 0 408 311, width=0.4\textwidth, clip]{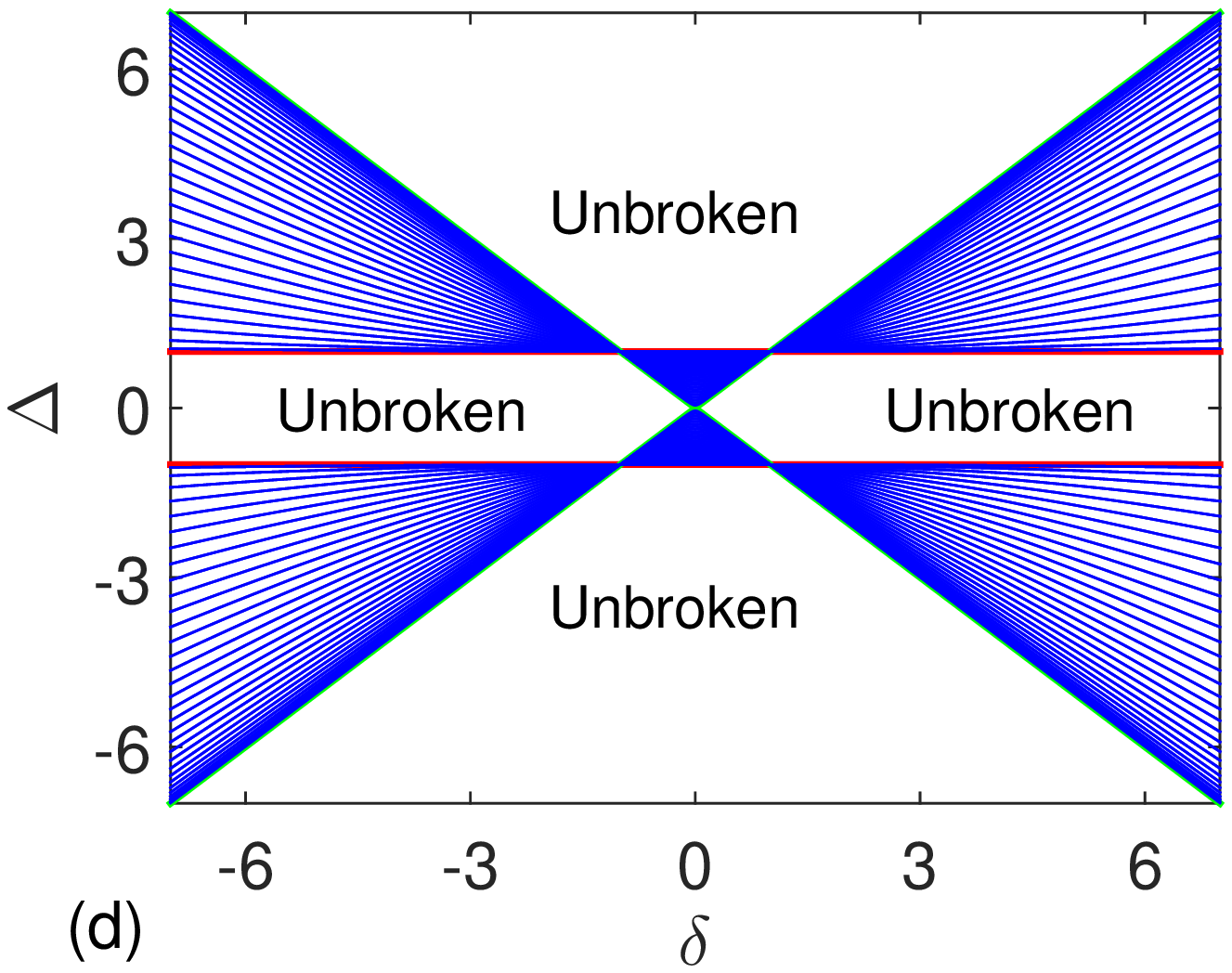}
\caption{(Color online) Plots of the phase diagrams for the systems with $%
N=4,$ $8,$ $24$ and $66$, respectively. The phase boundary denoted by red $%
\left( k_{c}=0\right) $ and green lines separates the $\protect\delta %
-\Delta $ plane into four regions. One can see that as $N$ increases, the
ELs covers the area between the boundary lines. Note that the broken region
does not include the origin, which corresponds to a degenerate point rather
than an EP.}
\label{fig3}
\end{figure*}
Before solving the Hamiltonian, we compare the Hamiltonian (\ref{H_SSH}) and
the Hamiltonian in Ref. \cite{HWH}. We note that the existence of the real
staggered potential $\lambda $ breaks the $\mathcal{PT}$ symmetry of the
system and therefore leads to complex energies. The corresponding
eigenstates do not possess any obvious symmetries except translation
symmetry. It strongly implies that the anti-linear symmetry of the
eigenstates is a key point in obtaining the full real spectrum. To obtain
the spectrum and corresponding eigenstates, we take the Fourier
transformation based on the translational symmetry,
\begin{eqnarray}
A_{k}^{\dagger } &=&\frac{1}{\sqrt{N}}\sum_{j}e^{ikj}c_{2j-1}^{\dagger }, \\
B_{k}^{\dagger } &=&\frac{1}{\sqrt{N}}\sum_{j}e^{ik\left( j+1/2\right)
}c_{2j}^{\dagger },
\end{eqnarray}%
satisfying%
\begin{equation}
T^{-1}A_{k}^{\dagger }T=e^{-ik}A_{k}^{\dagger },\text{ }T^{-1}B_{k}^{\dagger
}T=e^{-ik}B_{k}^{\dagger },
\end{equation}%
where $T$ is translational operator defined as $T^{-1}c_{j}^{\dagger
}T=c_{j+2}^{\dagger }$. The commutation relation $\left[ T,\mathcal{H}\right]
=0$ ensures that the Hamiltonian can be diagonalized in invariant subspace
spanned by the eigenvectors of operator $T$. Here $A_{k}$ and $B_{k}$ are
annihilation operators of fermions with $k=2n\pi /N$ $\left( n\in \left[ 1,N%
\right] \right) $ at odd and even sites, respectively. Then the Hamiltonian (%
\ref{H_SSH}) can be expressed as
\begin{eqnarray}
\mathcal{H} &=&\frac{1}{2}\sum_{k}\left[ \left( J_{0}\cos \frac{k}{2}%
-i\delta \sin \frac{k}{2}\right) A_{k}^{\dagger }B_{k}+\text{H.c.}\right]
\notag \\
&&+\left( i\Delta +\lambda \right) \sum_{k}\left( B_{k}^{\dagger
}B_{k}-A_{k}^{\dagger }A_{k}\right) \text{,}  \label{H_f}
\end{eqnarray}%
which is a bipartite lattice system consisting of two sub-lattices $A$ and $%
B $. By writing down the Hamiltonian (\ref{H_f}) in the Nambu representation
with the basis of $\left( A_{k}^{\dagger },B_{k}^{\dagger }\right) $, we have%
\begin{equation}
\mathcal{H}=\sum_{k}\left( A_{k}^{\dagger },B_{k}^{\dagger }\right) \mathcal{%
H}_{k}\left(
\begin{array}{c}
A_{k} \\
B_{k}%
\end{array}%
\right) ,
\end{equation}%
where the core matrix is%
\begin{equation}
\mathcal{H}_{k}=\left(
\begin{array}{cc}
-i\Delta -\lambda & J_{0}\cos \frac{k}{2}-i\delta \sin \frac{k}{2} \\
J_{0}\cos \frac{k}{2}+i\delta \sin \frac{k}{2} & i\Delta +\lambda%
\end{array}%
\right) .  \label{hk}
\end{equation}%
Note that $\mathcal{H}_{k}$ is similar to Eq. (\ref{H}), which is critical
to establish the classical correspondence of topological invariant. Based on
the fact that $\left[ \mathcal{H}_{k},\mathcal{H}_{k^{\prime }}\right] =0$,
this Hamiltonian can be diagonalized via introducing canonical operators,
\begin{eqnarray}
\alpha _{k} &=&u_{k}^{-}A_{k}+\nu _{k}^{-}B_{k},\text{ }\overline{\alpha }%
_{k}=u_{k}^{+}A_{k}^{\dagger }+\nu _{k}^{+}B_{k}^{\dagger }, \\
\beta _{k} &=&\zeta _{k}^{-}A_{k}+\xi _{k}^{-}B_{k},\text{ }\overline{\beta }%
_{k}=\zeta _{k}^{+}A_{k}^{\dagger }+\xi _{k}^{+}B_{k}^{\dagger },
\end{eqnarray}%
where the coefficients are%
\begin{eqnarray}
u_{k}^{\pm } &=&\frac{J_{0}\cos \left( k/2\right) \mp i\delta \sin \left(
k/2\right) }{\sqrt{2\varepsilon _{k}\left( \varepsilon _{k}+i\Delta +\lambda
\right) }}, \\
\text{ }v_{k}^{\pm } &=&\frac{\varepsilon _{k}+i\Delta +\lambda }{\sqrt{%
2\varepsilon _{k}\left( \varepsilon _{k}+i\Delta +\lambda \right) }}, \\
\zeta _{k}^{\pm } &=&\frac{J_{0}\cos \left( k/2\right) \mp i\delta \sin
\left( k/2\right) }{\sqrt{2\varepsilon _{k}\left( \varepsilon _{k}-i\Delta
-\lambda \right) }}, \\
\text{ }\xi _{k}^{\pm } &=&\frac{-\varepsilon _{k}+i\Delta +\lambda }{\sqrt{%
2\varepsilon _{k}\left( \varepsilon _{k}-i\Delta -\lambda \right) }}.
\end{eqnarray}%
Obviously, the complex modes $\left\{ \alpha _{k},\overline{\alpha }%
_{k},\beta _{k},\overline{\beta }_{k}\right\} $ satisfy the canonical
commutation relations%
\begin{eqnarray}
\left\{ \overline{\alpha }_{k},\alpha _{k^{\prime }}\right\} &=&\left\{
\overline{\beta }_{k},\beta _{k^{\prime }}\right\} =\delta _{kk^{\prime }},
\\
\left\{ \overline{\alpha }_{k},\overline{\alpha }_{k^{\prime }}\right\}
&=&\left\{ \overline{\beta }_{k},\overline{\beta }_{k^{\prime }}\right\} =0,
\\
\left\{ \overline{\alpha }_{k},\beta _{k^{\prime }}\right\} &=&\left\{
\overline{\beta }_{k},\alpha _{k^{\prime }}\right\} =0, \\
\left\{ \alpha _{k},\alpha _{k^{\prime }}\right\} &=&\left\{ \beta
_{k},\beta _{k^{\prime }}\right\} =0.
\end{eqnarray}%
These relations enable the establishment of the biorthogonal bases thereby
diagonalizing the Hamiltonian
\begin{equation}
\mathcal{H}=\sum_{k}\varepsilon _{k}\left( \overline{\alpha }_{k}\alpha _{k}-%
\overline{\beta }_{k}\beta _{k}\right) ,
\end{equation}%
here the single-particle spectrum in each subspace is
\begin{equation}
\varepsilon _{k}=\sqrt{J_{0}^{2}\cos ^{2}\left( k/2\right) +\delta ^{2}\sin
^{2}\left( k/2\right) +\left( i\Delta +\lambda \right) ^{2}}.
\end{equation}%
Note that the Hamiltonian $\mathcal{H}$ is still non-Hermitian owing to the
fact that $\overline{\alpha }_{k}\left( \overline{\beta }_{k}\right) \neq
\alpha _{k}^{\dagger }\left( \beta _{k}^{\dagger }\right) $. Accordingly,
the eigenstates of $\mathcal{H}$ can be written in the form
\begin{equation}
\prod\limits_{\left\{ k\right\} }\overline{\eta }_{k}\left\vert
Vac\right\rangle ,
\end{equation}%
where $\overline{\eta }_{k}$ represents either $\overline{\alpha }_{k}$ or $%
\overline{\beta }_{k}$. This constructs the biorthogonal set associated with
the eigenstates
\begin{equation}
\left\langle Vac\right\vert \prod\limits_{\left\{ k\right\} }\eta _{k}
\end{equation}%
of the Hamiltonian $\mathcal{H}^{\dag }$, where $\left\vert Vac\right\rangle
$ is the vacuum state of the fermion $c_{j}$. Next we move focus on the
eigenstate $\left\vert GS\right\rangle $, which can be given as
\begin{equation}
\left\vert GS\right\rangle =\prod\limits_{k}\overline{\beta }_{k}\left\vert
Vac\right\rangle ,
\end{equation}%
and the corresponding birothogonal eigenstate of $\mathcal{H}^{\dag }$ is
\begin{equation}
\left\vert \overline{GS}\right\rangle =\prod\limits_{k}\beta _{k}^{\dag
}\left\vert Vac\right\rangle .
\end{equation}%
The corresponding eigenenergy $-\sum_{k}\varepsilon _{k}$ is complex due to
the existence of the complex on-site staggered potentials. As a reminder,
these complex potentials also break the $\mathcal{PT}$ symmetry of the
system. We would like to point out that the system can have full real
spectrum in the absence of $\lambda $, which depends on whether the
condition of $\left[ J_{0}^{2}\cos ^{2}\left( k/2\right) +\delta ^{2}\sin
^{2}\left( k/2\right) -\Delta ^{2}\right] >0$ holds up. In the unbroken
region, i.e, Im($\varepsilon _{k}$)$=0$, $\left\vert GS\right\rangle $ is
the ground state of the system while the system possesses two gapped bands
\cite{HWH}. The alteration of the system parameters gives rise to the
variation of the eigenstate symmetry. We are not concerned with whether the
eigenstate $\left\vert GS\right\rangle $ is the ground state of the system.
In this paper, we focus on the dynamical property of the $\left\vert
GS\right\rangle $ in the following.

\subsection{Phase diagram and EL}

Now we turn to study the EPs of the system based on the previous solution.
The EPs except the point $\left\{ \Delta =\lambda =\delta =0,\text{ }%
k_{c}=\pi \right\} $, which corresponds to the degeneracy point rather than
EP, can be identified via the condition $\varepsilon _{k}=0$. It is clear
that when $\lambda =0$, any one of the momentum $k$ satisfies
\begin{equation}
J_{0}^{2}\cos ^{2}\left( k/2\right) +\delta ^{2}\sin ^{2}\left( k/2\right)
-\Delta ^{2}=0,
\end{equation}%
admitting an exceptional curve in the $\delta -\Delta $ plane, at which the
coefficients $u_{k}^{\pm }$, $v_{k}^{\pm }$, $\zeta _{k}^{\pm }$ and $\xi
_{k}^{\pm }$ associated with corresponding canonical operators diverges.
This feature is in agreement with that in the Ref. \cite{ZXZ}. In addition,
the eigenstate of the Nambu expression of $\mathcal{H}_{k}$ coalesces
supporting another feature of EP. In Fig. \ref{fig3}, we plot the phase
diagrams of the system. It shows that the ELs lie in the $\delta -\Delta $
plane. Each EL corresponds to two critical momentums $\pm k_{c}$, which is
crucial to demonstrate the topological property of $\left\vert
GS\right\rangle $.

It is worth pointing out that when $\Delta =0$, the system reduces to the
Hermitian RM model. The two bands in this kind of systems with even sites
touch each other at the point $\left\{ \lambda =\delta =0,\text{ }k_{c}=\pi
\right\} $. However, the EL can exist in the finite size of the system. In
the thermodynamic limit $N\rightarrow \infty $, the phase boundary can be
determined by the two straight lines corresponding to $k_{c}=0 $, $\pi $,
which is different from the dog-leg line in Ref. \cite{ZXZ}. We plot these
two lines in Fig. \ref{fig3} with a red and a green lines, respectively.
Moreover, in previous works \cite{TonyPRL}, the topology of the EP can be
characterized through the Zak phase which is associated with the
bulk-boundary correspondence. The introducing of the Zak phase requires an
infinite system with continuous $k$. This approach is not applicable to the
study of finite non-Hermitian systems. In the following discussion, we
remove this constraint and investigate the topology of the EP through the
variation of the system parameters in the parameter space.

\subsection{Topological invariant and classical correspondence}

\begin{figure*}[tbp]
\centering
\includegraphics[bb=100 230 480 530, width=0.3\textwidth, clip]{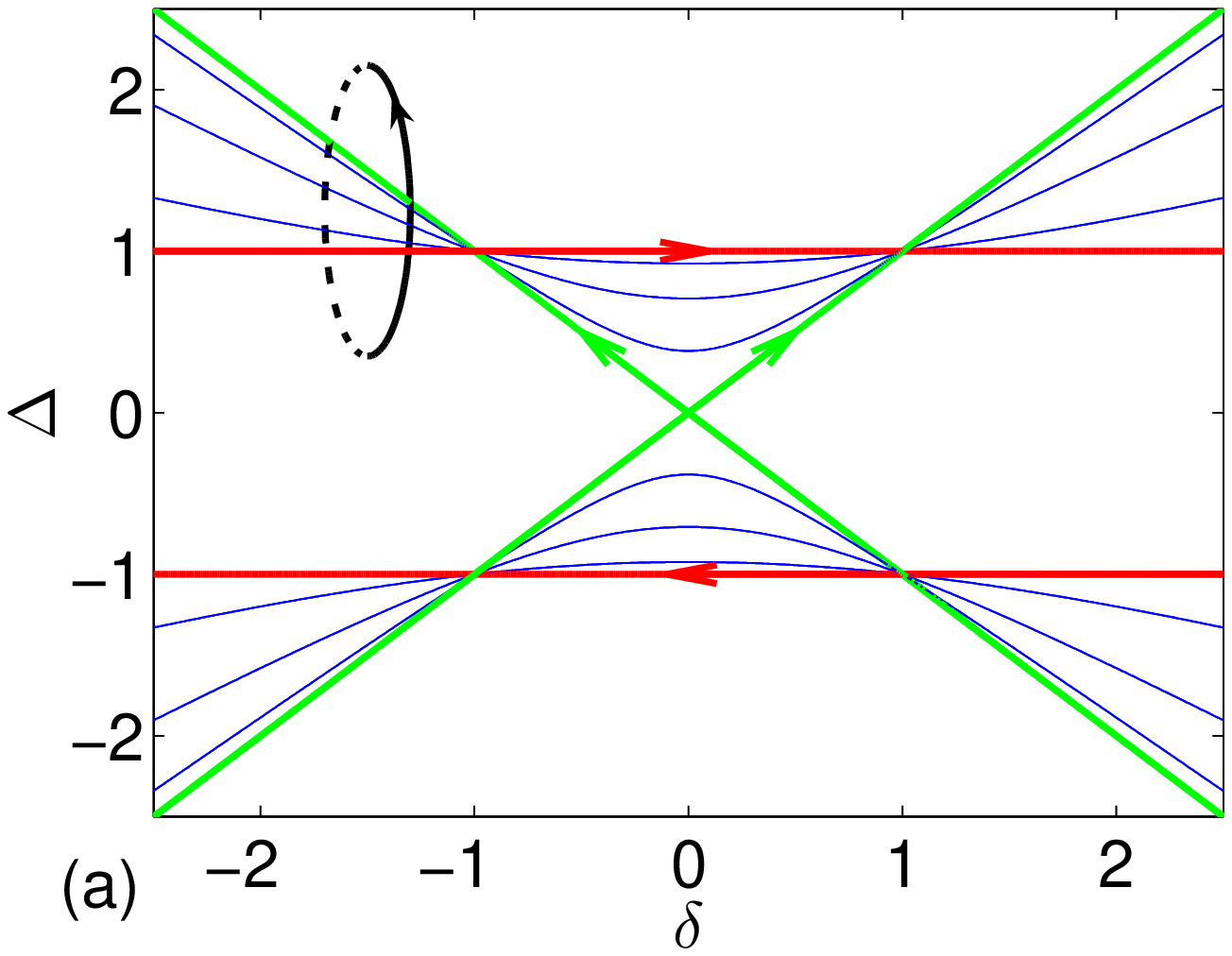} %
\includegraphics[bb=100 230 480 530, width=0.3\textwidth, clip]{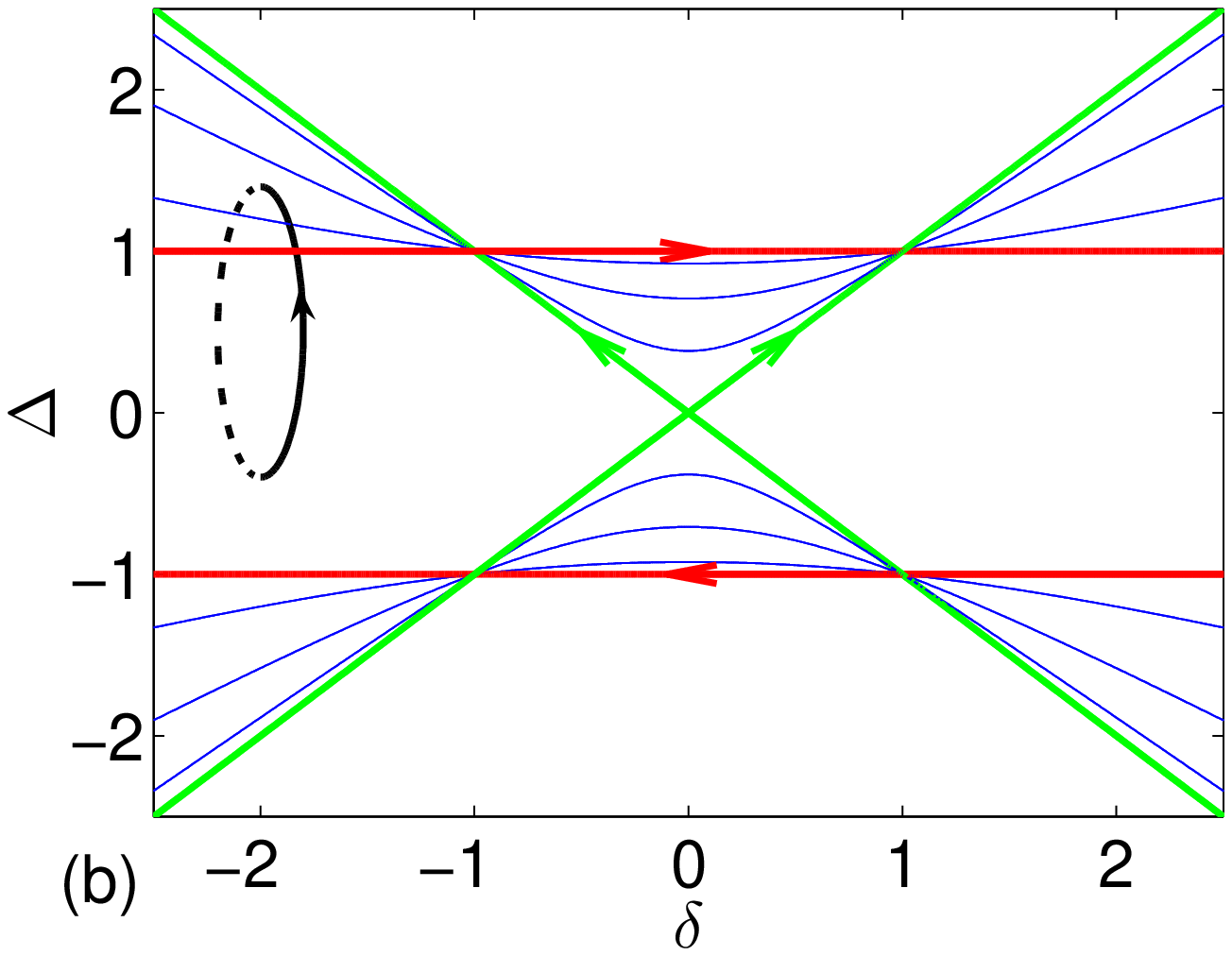} %
\includegraphics[bb=100 230 480 530, width=0.3\textwidth, clip]{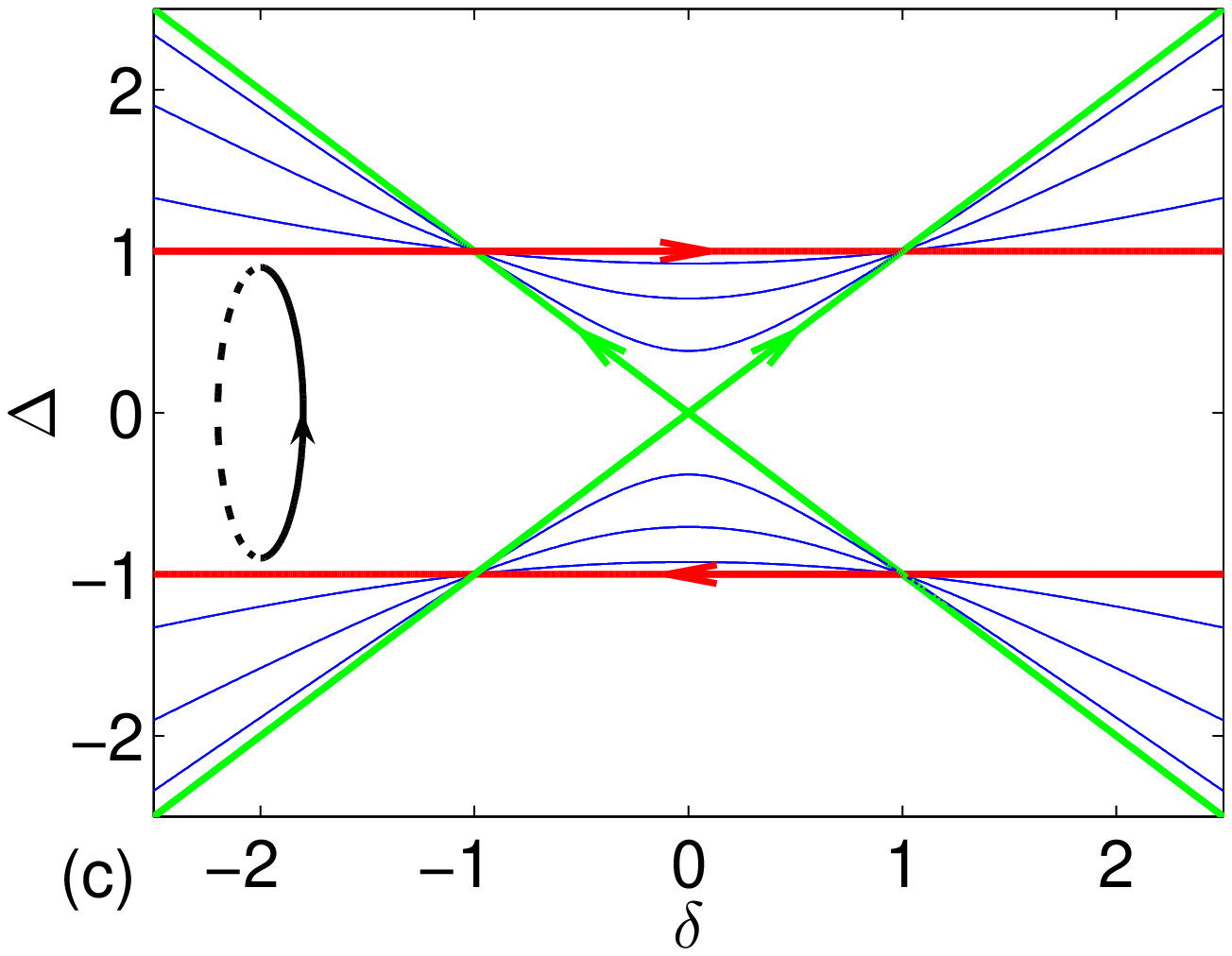} %
\includegraphics[bb=100 230 480 530, width=0.3\textwidth, clip]{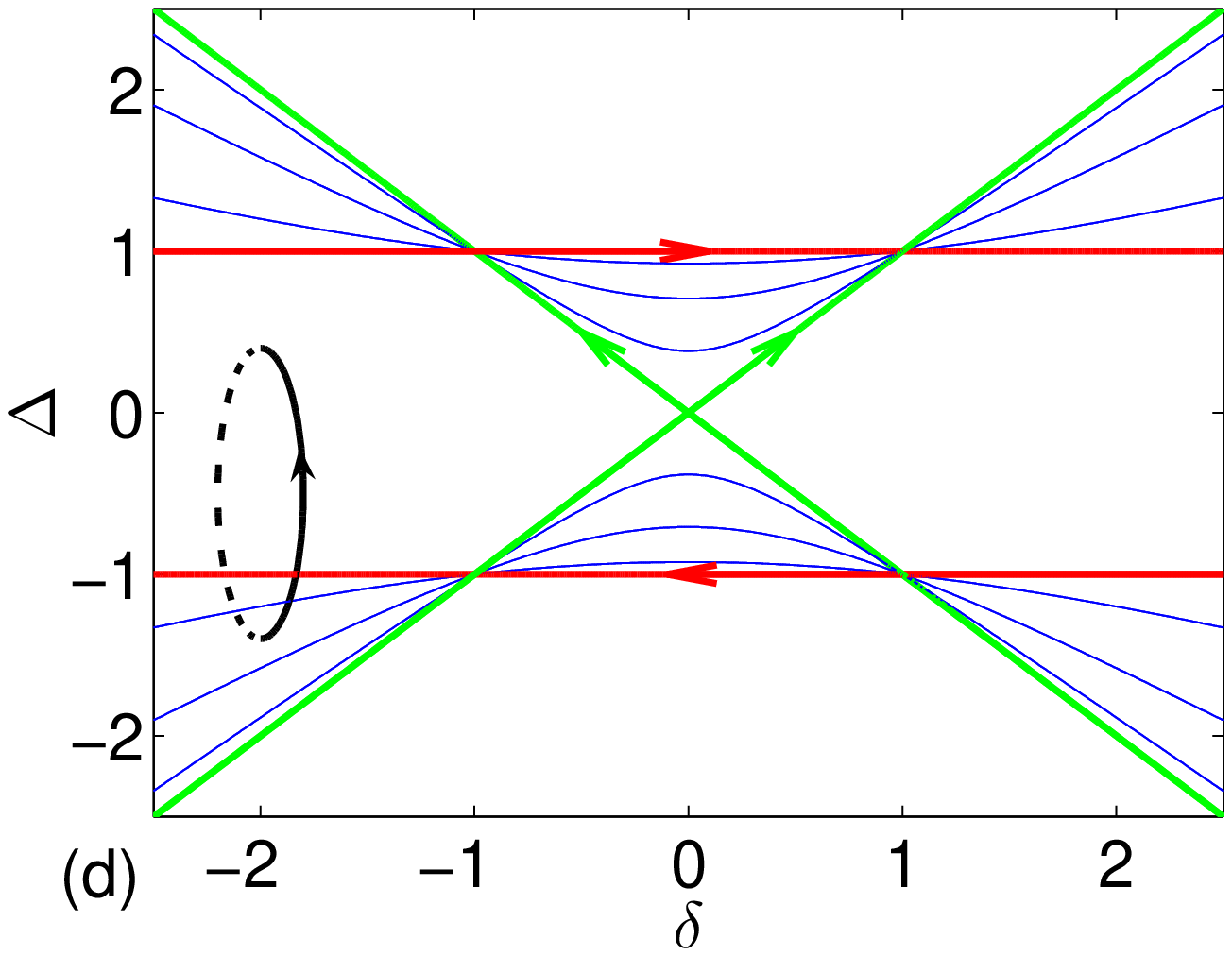} %
\includegraphics[bb=100 230 480 530, width=0.3\textwidth, clip]{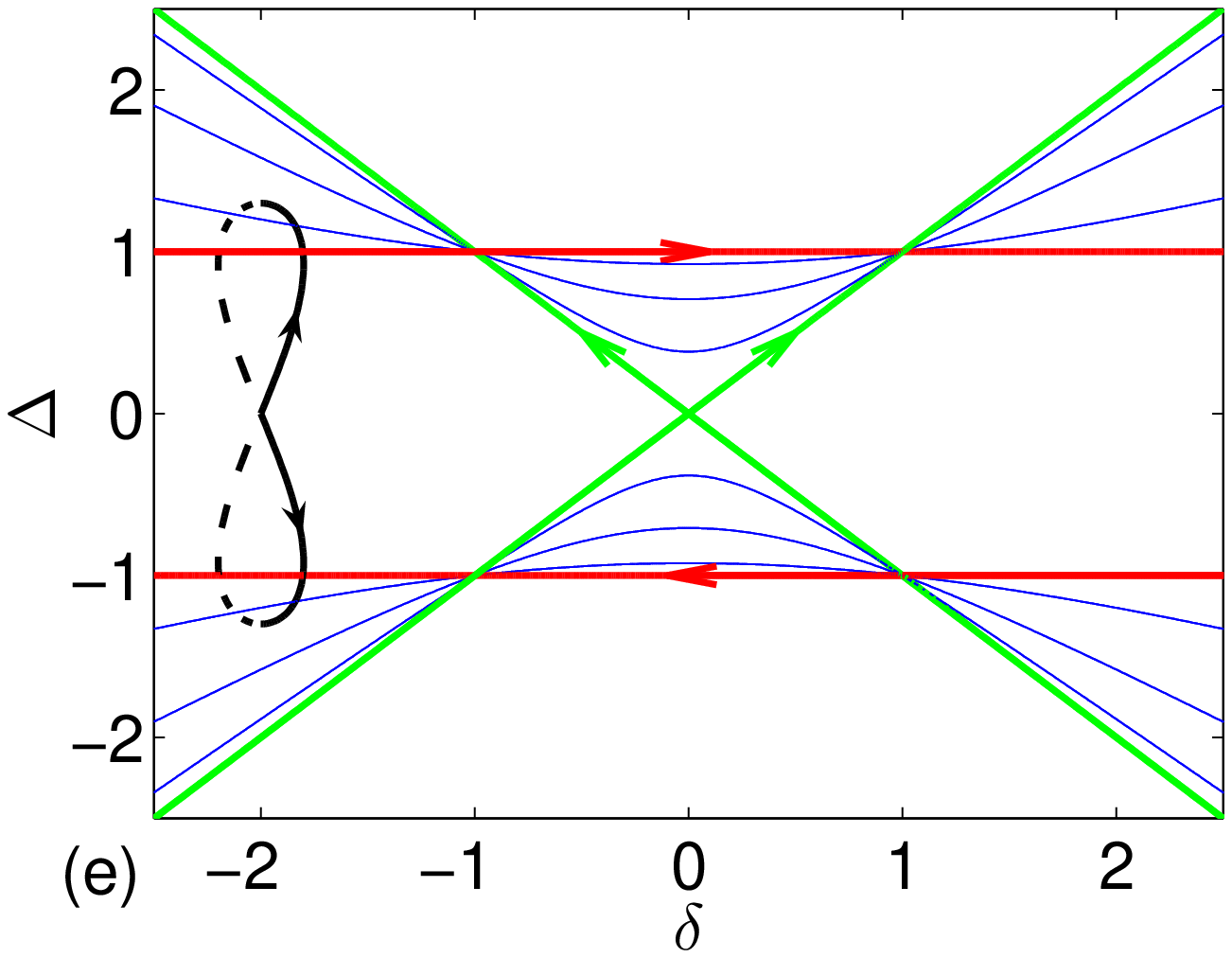} %
\includegraphics[bb=100 230 480 530, width=0.3\textwidth, clip]{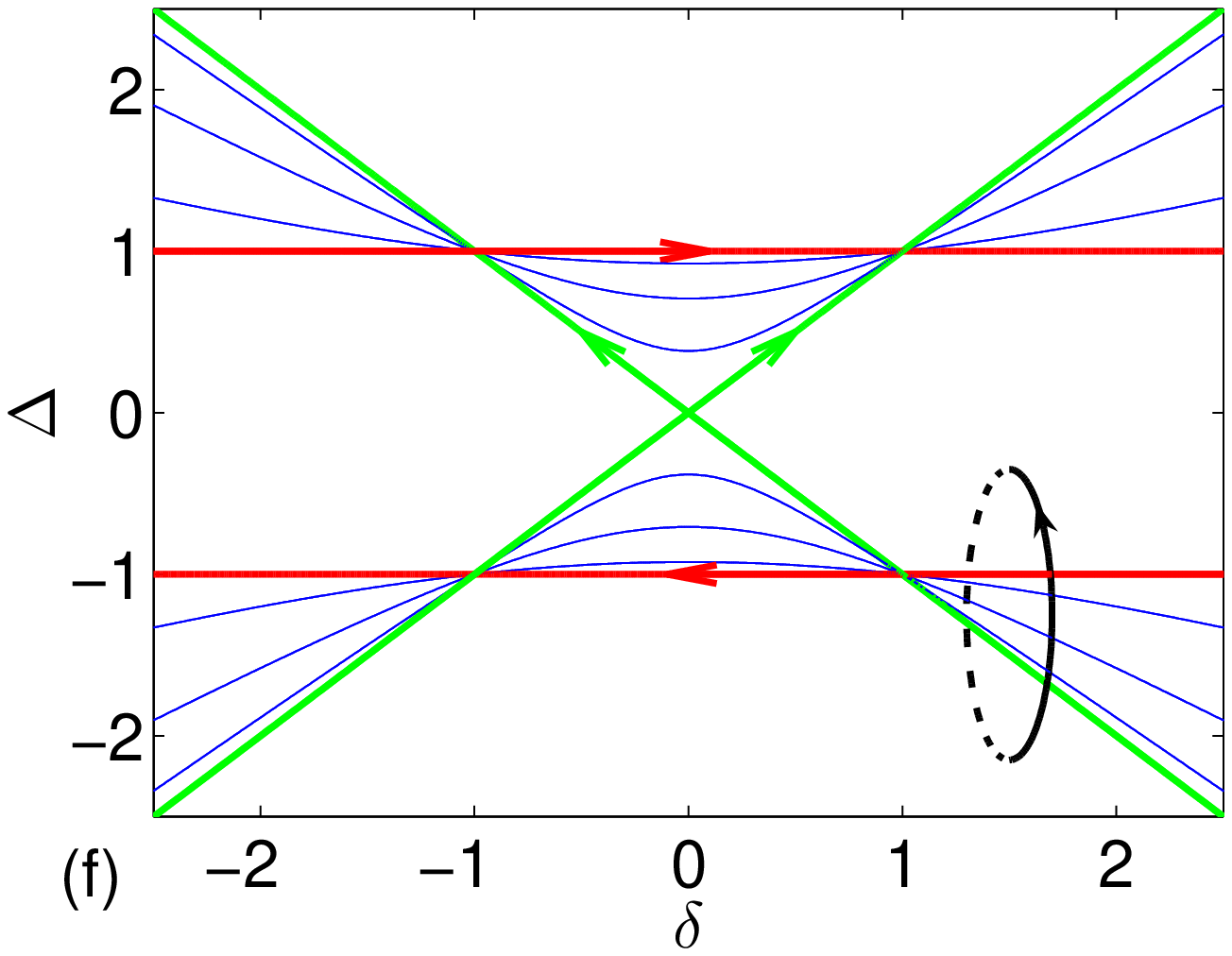}
\caption{(Color online) Schematic illustration of different paths
corresponding to the different values of Berry phases in the phase diagram
with $N=8$. The black curve represents the closed evolution path and the
other lines are the same as those in Fig. \protect\ref{fig3}. The Berry
phases are: (a) $0$, (b) $-\protect\pi/2 $, (c) $0$, (d) $\protect\pi/2 $,
(e) $-\protect\pi $, (f) $\protect\pi $.}
\label{fig4}
\end{figure*}
To demonstrate the topology of the system, according to the main conclusion
in Sec. II, we assume that the Hamiltonian (\ref{H_SSH}) is a periodic
function of the system parameters $\left\{ \lambda ,\text{ }\Delta ,\text{ }%
\delta \right\} $ that vary with time $t$, and $J_{0}$ is taken as a
constant. The corresponding Berry connection $\overrightarrow{\mathcal{A}}%
=\left( \mathcal{A}_{\lambda },\text{ }\mathcal{A}_{\Delta },\text{ }%
\mathcal{A}_{\delta }\right) $ of the eigenstate $\left\vert GS\left(
t\right) \right\rangle $ can be expressed as
\begin{equation}
\mathcal{A}_{g}=i\left\langle \overline{GS}\left( t\right) \right\vert \frac{%
\partial }{\partial g}\left\vert GS\left( t\right) \right\rangle ,
\end{equation}%
where $\left\vert GS\left( t\right) \right\rangle $ represents the
instantaneous eigenstate of the Hamiltonian $H\left( \lambda \left( t\right)
,\Delta \left( t\right) ,\delta \left( t\right) \right) $, and $g=\lambda $,
$\Delta $, $\delta $ denote the three directions in the parameter space. The
Berry connection can be explicitly expressed as%
\begin{equation}
\overrightarrow{\mathcal{A}}=\sum_{k}\overrightarrow{\mathcal{A}}_{k},
\end{equation}%
where $\overrightarrow{\mathcal{A}}_{k}=\left( \mathcal{A}_{\lambda ,k},%
\text{ }\mathcal{A}_{\delta ,k},\text{ }\mathcal{A}_{\Delta ,k}\right) $, and%
\begin{equation}
\mathcal{A}_{g,k}=i\left\langle Vac\right\vert \beta _{k}\frac{\partial }{%
\partial g}\overline{\beta }_{k}\left\vert Vac\right\rangle
\end{equation}%
represents the Berry connection of the single-particle eigenstate. After
some straightforward algebras, we have%
\begin{equation}
\overrightarrow{\mathcal{A}}=\overrightarrow{\mathcal{A}}_{0}+%
\overrightarrow{\mathcal{A}}_{\pi },
\end{equation}%
where%
\begin{eqnarray}
\overrightarrow{\mathcal{A}}_{0} &=&\left( J_{0},\text{ }iJ_{0},\text{ }%
0\right) /\mathcal{D}_{0}, \\
\overrightarrow{\mathcal{A}}_{\pi } &=&\left( \delta ,\text{ }i\delta ,\text{
}-\lambda -i\Delta \right) /\mathcal{D}_{\pi },
\end{eqnarray}
with $\mathcal{D}_{0}=2\varepsilon _{0}^{2}$ and $\mathcal{D}_{\pi
}=2\varepsilon _{\pi }^{2}$. Here we can see that the Berry connection
consists of two components, the contribution of which comes from the $k=0$, $%
\pi $, respectively. All the other $k$ contributes nothing to the total
Berry connection owing to the fact that $\overrightarrow{\mathcal{A}}_{k}=-%
\overrightarrow{\mathcal{A}}_{2\pi -k}$. Note that we choose different
gauges so that the single-particle eigenstates $\overline{\beta }%
_{0}\left\vert Vac\right\rangle $, $\overline{\beta }_{k}\left\vert
Vac\right\rangle $ are smooth and single valued everywhere \cite{Xiao}.
According to the classical correspondence in Eq. (\ref{eqv}), the boundary
lines denoted by red and green color in Fig. \ref{fig3} can act as four
vortex filaments, the directions of which are sketched in Fig. \ref{fig4}.
Therefore the Berry connection $\overrightarrow{\mathcal{A}}$ can be further
expressed as%
\begin{eqnarray}
\overrightarrow{\mathcal{A}}_{0} &=&\overrightarrow{A}_{0}+i\nabla \varphi
_{0},  \label{eqv1} \\
\overrightarrow{\mathcal{A}}_{\pi } &=&\overrightarrow{A}_{\pi }+i\nabla
\varphi _{\pi },  \label{eqv2}
\end{eqnarray}%
where
\begin{equation}
\overrightarrow{A}_{0}=\sum_{\sigma =\pm }\sigma \overrightarrow{A}%
_{0,\sigma },\text{ }\overrightarrow{A}_{\pi }=\sum_{\sigma =\pm }\sigma
\overrightarrow{A}_{\pi ,\sigma },
\end{equation}%
and the scalar potentials are%
\begin{eqnarray}
\varphi _{0} &=&\frac{1}{8}\ln \frac{D_{0,-}}{D_{0,+}}, \\
\varphi _{\pi 0} &=&\frac{1}{8}\ln \frac{D_{\pi ,-}}{D_{\pi ,+}}.
\end{eqnarray}%
with
\begin{eqnarray}
D_{0,\pm } &=&\left( \Delta \pm J_{0}\right) ^{2}+\lambda ^{2}, \\
D_{\pi ,\pm } &=&\left( \Delta \pm \delta \right) ^{2}+\lambda ^{2}.
\end{eqnarray}%
Moreover, the four curl fields with the form of
\begin{eqnarray}
\overrightarrow{A}_{0,\pm } &=&\left( -\Delta \pm J_{0},\text{ }\lambda ,%
\text{ }0\right) /4D_{0,\mp }, \\
\overrightarrow{A}_{\pi ,\pm } &=&\left( -\Delta \pm \delta ,\text{ }\lambda
,\text{ }\mp \lambda \right) /4D_{\pi ,\mp },
\end{eqnarray}%
are induced by vortex filaments of $\Delta =\pm J_{0}$, $\Delta =\pm \delta $
in $\Delta -\delta $ plane, respectively. In the context of classical
electromagnetics, such the vortex filaments can be deemed as the magnetic
field lines induced by solenoid. On the other hand, we consider the
non-boundary ELs (blue hyperbolas in Fig. \ref{fig3})that have nothing to do
with the total Berry connection. It seems that each blue line is generated
by the two hyperbolic solenoids with opposite current directions. However,
the direct calculation shows that such a correspondence is incorrect. One
cannot establish the connection between $\overrightarrow{\mathcal{A}}_{k}$
and $\overrightarrow{A}_{k}$ like Eqs. (\ref{eqv1}) and (\ref{eqv2}). In
other words, the blue lines are not topological. This can also be understood
in another way: For the momentum $k\neq 0$, $\pi $, the core matrix (\ref{hk}%
) can be decomposed into three Pauli operators as
\begin{equation}
\mathcal{H}_{k}=\overrightarrow{B}\cdot \overrightarrow{\sigma },
\end{equation}%
where $\overrightarrow{B}=\left( J_{0}\cos \frac{k}{2},\text{ }\delta \sin
\frac{k}{2},\text{ }-\lambda -i\Delta \right) $. For the current form of
field $\overrightarrow{B}$, a performance of a spin rotation reduces the
system to containing two Pauli operators. However, such a spin rotation
operator must be dependent on the system parameters $\left\{ \lambda ,\text{
}\Delta ,\text{ }\delta \right\} $. This also indicates that there does not
exist a parameter-independent chiral operator $C$ satisfying $\left\{ C,%
\mathcal{H}_{k}\right\}=0 $. Therefore, the main conclusion in Sec. II
cannot be applied to the blue lines. In this point, not all the ELs can have
classical correspondence.

Moreover, the corresponding Berry phase undergoing a closed path in the
parameter space can be further expressed as $\gamma _{B}=\int_{C}%
\overrightarrow{\mathcal{A}}\cdot d\overrightarrow{l}=\int_{C}%
\overrightarrow{A}\cdot d\overrightarrow{l}$, which is identical to the
magnetic flux penetrating the corresponding enclosed surface. In this point
of view, the topology of the EL characterized by the Berry phase depends on
the path of the closed curve. If the rotation direction of the curve judged
by right handed screw rule is the same as the direction of red or green
line, there is a $\pi /2$ contribution to Berry phase. Conversely, there is
a $-\pi /2$ contribution to Berry phase. Furthermore, the blue lines does
not contribute to the Berry phase because there are two opposing magnetic
fluxes. As a result, the Berry phase is the sum of the contribution of red
and green lines that penetrate the enclosed surface. We sketch several paths
in Fig. \ref{fig4}, which corresponds to the different values of Berry
phases (topological quantum numbers). Here are only a few simple examples in
which the values of the Berry phase are taken between $-\pi $ and $\pi $.
Through changing the form of the curve, the Berry phase can be any integer
multiples of $\pi /2$. We would like to present here that the non-zero Berry
curvature can be served as a dynamical signature to identify the existence
of the EP in non-Hermitian systems.

For the finite system, the evolution path can pass through the region
without ELs between red and green boundary lines. The topology of the EP can
be characterized by the Berry phase in the parameter space rather than the
Zak phase in the $k$ space of the infinite system. This provides an
alternative way to describe the topology of the EP in finite non-Hermitian
many-particle system. For the infinite system, the Berry phase is an integer
multiples of $\pi $ since the existence of the EP lines prevent the
selection of the closed path surrounding a single boundary line.

\section{Summary and discussion}

In summary, we have studied systematically the topology of the ELs in the
parameter space through a simple non-Hermitian matrix. Based on the exact
solution, the EL can be equivalent to a vortex filament in the classical
physics. The curl field induced by the vortex filament is connected to the
Berry connection of the non-Hermitian system through a gauge transformation.
We apply this result to the core matrix of the non-Hermitian RM model and
then find that the topological properties can be characterized by the Berry
phase in the parameter space rather than Zak phase accumulated by an
eigenstate during its parallel transport through the whole Brillouin zone.
This findings provides an alternative way to investigate the topology of the
EP in an experimentally accessible finite non-Hermitian lattice model with a
discrete momentum space. Furthermore, based on the exact solution, we have
shown that the boundary ELs associated with critical momenta $k_{c}=0$, $\pi
$ can act as magnetic field lines induced by the infinite solenoids with
infinitesimal diameter. For the non-boundary ELs, each line corresponds to
two critical momenta $k_{c}$, $2\pi -k_{c}$ $\left( k_{c}\neq 0,\text{ }\pi
\right) $, so all the non-boundary ELs make no contribution. Therefore, in
this context, the Berry connection in the parameter space is equivalent to
the vector potential generated by the four boundary ELs. From this
perspective, the Berry phase is limited to an integer multiples of $\pi /2$,
which is dependent on the magnetic flux of the area enclosed by trajectory. %
In addition, we would like to point that each boundary EL
can have a direction, which can be determined by the corresponding Berry
curvature. The non-zero Berry curvature can be deemed as a dynamical
signature to identify the existence of the EP in the non-Hermitian systems.

\acknowledgments We thank X. Q. Li for helpful discussions and comments.
This work is supported by the National Natural Science Foundation of China
(Grants No. 11705127, No. 11505126, and No. 11374163). G. Zhang and X. Z.
Zhang are also supported by PhD research startup foundation of Tianjin
Normal University under Grants No. 52XB1415 and No. 52XB1608.

\end{document}